\documentclass[aps,prb,superscriptaddress,twocolumn]{revtex4-2}

\usepackage{graphicx}
\usepackage{amsfonts,amsmath,amssymb,amsthm}
\usepackage{epstopdf}
\usepackage{upgreek,xspace}
\usepackage{chngcntr}
\usepackage{hyperref}
\hypersetup{colorlinks,allcolors=blue}
\usepackage[version=3]{mhchem} 
\usepackage{soul}

\usepackage{setspace}

\newcommand{\be}{\begin{equation}}
\newcommand{\ee}{\end{equation}}

\newcommand{\Tg}{$T_\mathrm{G}$}
\newcommand{\Po}{$P_\mathrm{O_2}$}
\newcommand{\ALO}{Al$_2$O$_3$}

\begin{document}

\title{High temperature diffusion enabled epitaxy of the Ti-O system}

\author{Jeong~Rae~Kim}\thanks{These authors contributed equally}
\affiliation{Department of Applied Physics and Materials Science, California Institute of Technology, Pasadena, California 91125, USA.}
\affiliation{Institute for Quantum Information and Matter, California Institute of Technology, Pasadena, California 91125, USA.}

\author{Sandra~Glotzer}\thanks{These authors contributed equally}
\affiliation{Department of Applied Physics and Materials Science, California Institute of Technology, Pasadena, California 91125, USA.}
\affiliation{Institute for Quantum Information and Matter, California Institute of Technology, Pasadena, California 91125, USA.}

\author{Adrian~Llanos}
\affiliation{Department of Applied Physics and Materials Science, California Institute of Technology, Pasadena, California 91125, USA.}
\affiliation{Institute for Quantum Information and Matter, California Institute of Technology, Pasadena, California 91125, USA.}

\author{Salva~Salmani-Rezaie}
\affiliation{Department of Materials Science and Engineering, The Ohio State University, Columbus, Ohio 43210, USA}

\author{Joseph~Falson}
\email{falson@caltech.edu}
\affiliation{Department of Applied Physics and Materials Science, California Institute of Technology, Pasadena, California 91125, USA.}
\affiliation{Institute for Quantum Information and Matter, California Institute of Technology, Pasadena, California 91125, USA.}


\begin{abstract}
High temperatures promote kinetic processes which can drive crystal synthesis towards ideal thermodynamic conditions, thereby realizing samples of superior quality. While accessing very high temperatures in thin-film epitaxy is becoming increasingly accessible through laser-based heating methods, demonstrations of such utility are still emerging. Here we realize a novel self-regulated growth mode in the Ti-O system by relying on thermally activated diffusion of oxygen from an oxide substrate. We demonstrate oxidation selectivity of single phase films with superior crystallinity to conventional approaches as evidenced by structural and electronic measurements. The diffusion-enabled mode is potentially of wide use in the growth of transition metal oxides, opening up new opportunities for ultra-high purity epitaxial platforms based on \textit{d}-orbital systems.
\end{abstract}

\flushbottom
\maketitle

\section*{Introduction}
There is a dearth of ultra-high-quality crystals as thin films compared to bulk samples. This stems from the unique kinetic and thermodynamic constraints when growing compounds on a surface. The growth process generally occurs well below the melting temperature of the compound, which places the process far away from perfect diffusion models and therefore necessitates the accurate composition control to suppress the formation of unwanted phases. While ramping up the growth temperature can in theory promote kinetic processes which assist in forming high-quality crystals, there are limitations on how far it can be increased without destructively affecting the thin film structures. Examples include melting/evaporation of the material being grown, chemical reactions between adjacent layers, and diffusion of elements across heterointerfaces.

Since an abrupt interface of two heterogeneous materials can be seen as a low-entropy state, increasing temperature will also kinetically activate diffusion phenomena, which will act as a fundamental compensatory action of the system \cite{fultzPhaseTransitionsMaterials2020}. 
Undesired diffusion often results in rough heterointerfaces that lead to the deterioration of sample quality~\cite{nakagawaWhyInterfacesCannot2006}. In the case of oxide heterostructures, oxygen diffusion can also cause unintended oxidation or reduction of a metal species~\cite{scheidererTailoringMaterialsMottronics2018,dasDeterministicInfluenceSubstrateInduced2021}, which, in addition to causing atomic scale disorder, will hinder the ability to achieve phase-pure films. In the field of quantum materials, where energy scales of competing electronic and magnetic phenomena can be a fraction of an electron volt, even single atomic defects can have a detrimental impact on the variety and clarity of resolvable phases \cite{hwangEmergentPhenomenaOxide2012}. 

In recent years, cases have been identified where oxygen diffusion is advantageous in the fabrication of oxide heterostructures. It has assisted in the formation of a two-dimensional electron gas of SrTiO$_3$~\cite{posadasScavengingOxygenSrTiO32017,chenHighmobilityTwodimensionalElectron2013} and solid-state reduction of superconducting infinite-layer nickelates~\cite{weiSuperconductingNd1xEuxNiO2Thin2023}. Additionally, it has been reported that oxygen diffusion from the substrate can oxidize the growing layer~\cite{rosenbergerEuropiumOxideGrowth2022,posadasFacileGrowthEpitaxial2022}, i.e., stabilize the oxide film. These circumstances motivate the use of ultra-high growth temperatures to explore the limits of diffusion and whether it can offer advantageous pathways to epitaxial film growth.

Here, we demonstrate a self-regulated epitaxy mechanism applicable to transition metal oxides using the Ti-O system at very high synthesis temperatures using molecular beam epitaxy (MBE). The novelty of our result is two-fold. First, we demonstrate the ability to synthesize these structures with oxidation state selectivity without oxidant being injected into the growth chamber. Second, the films grown at high temperature elicit highly reproducible structural and electronic properties across a wide range of growth parameters, pointing to the establishment of growth dynamics which closely resemble adsorption-controlled pathways. We attribute this growth mode to thermally-activated oxygen diffusion from the substrate acting as a source of oxidation of the titanium metal.

\begin{figure}[t]
  \centering
   \includegraphics[width=85mm]{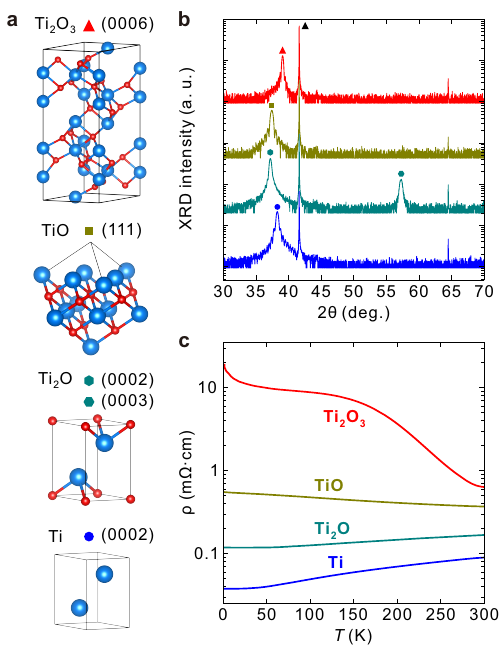}
   \caption{Overview of the Ti-O system. \textbf{a}, Crystal structures of Ti, Ti$_2$O, TiO, and Ti$_2$O$_3$. \textbf{b}, Out-of-plane X-ray diffraction (XRD) of the representative thin film samples. Blue, green, yellow, red, and black markers in \textbf{b} denote the XRD peaks of Ti, Ti$_2$O, TiO, Ti$_2$O$_3$, and Al$_2$O$_3$ substrate. \textbf{c}, Temperature dependent resistivity of the samples in \textbf{b}}
   \label{Fig1}
\end{figure}

\section*{Results and discussion}

\subsection*{The Ti-O thin film system}

We have extensively explored Al$_2$O$_3$ (0001) substrates as a platform for this growth mode due to their exceptional thermal stability and conductivity, strong ionic bonds and simple oxidation chemistry. To achieve very high temperatures, we use a CO$_2$ laser heating apparatus which effortlessly can take the substrate to temperatures exceeding 2000$^\circ$C \cite{andersonGrowthTiSapphire1997a,braunSituThermalPreparation2020a}. Ti as a transition metal can take numerous oxidation states, of which we will focus on 0+ (Ti), 1+ (Ti$_2$O), 2+ (TiO), 3+ (Ti$_2$O$_3$). These structures are shown in Fig.\ref{Fig1}\textbf{a}, with Ti, Ti$_2$O, and Ti$_2$O$_3$ taking hexagonal unit cells, while TiO is cubic (Table \ref{table}). The fully-oxidized end member TiO$_2$ is not the focus of this study. The corresponding out-of-plane diffraction for a set of phase-pure films is shown in Fig.\ref{Fig1}\textbf{b}, illustrating that both peak position and the diffracting crystal planes are unique for each structure. Utilizing more extensive peak searching in asymmetric diffraction conditions (Fig.\ref{SI_RSM}), we are able to conclusively identify which phases are stabilized in our growth process.

\begin{figure}[t]
  \centering
   \includegraphics[width=85mm]{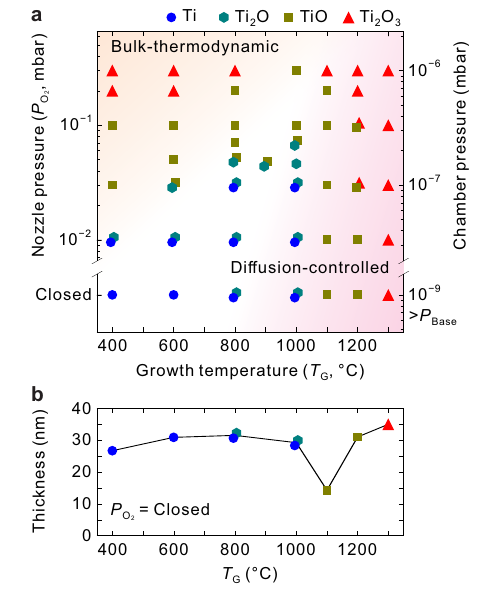}
   \caption{\textbf{a}, Ti-O growth phase diagram in the \Tg-\Po~parameter space. The shaded regions are intended to indicate the approximate bounds of the ``bulk-thermodynamic'' and ``diffusion-controlled'' growth regimes. \textbf{b}, Thicknesses of the Ti-O films grown without supplying oxygen, calculated by fitting X-ray diffraction and reflectivity.}
   \label{Fig2}
\end{figure}

Electrical transport also helps us understand the crystalline phase, as stripping electrons from Ti$^{0+}$ also progressively reduces the metallicity of the structure towards a strongly correlated metal (TiO)~\cite{hulmSuperconductivityTiONbO1972a,banusElectricalMagneticProperties1972}, and ultimately a correlated insulator (Ti$_2$O$_3$) \cite{uchidaChargeDynamicsThermally2008a,imadaMetalinsulatorTransitions1998b}. This is illustrated in Fig.\ref{Fig1}\textbf{c} where we have plotted the temperature-dependent electrical resistivity ($\rho$) of a set of phase-pure films. There have been extensive studies on binary titanium oxide thin films focusing on their correlated electron phases, and several studies have pointed out the difficulty of accurately controlling the titanium-oxygen composition ratio \cite{liSinglecrystallineEpitaxialTiO2021,yoshimatsuSuperconductivityTi4O7GTi3O52017a,yoshimatsuLargeAnisotropyConductivity2018,liObservationSuperconductivityStructureselected2018a,yoshimatsuMetallicGroundStates2020a,shvetsSuppressionMetalinsulatorTransition2020,yoshimatsuEvidenceLatticeDeformation2022a}. For example, the Ti$_2$O phase has been reported to a very limited extent even in bulk synthesis studies~\cite{holmbergDisorderOrderSolid1962,fanStructureTransportProperties2019}, and the TiO phase is known to accommodate a large compositional variation, ranging from TiO$_{0.8}$ to TiO$_{1.2}$ (or a wider range), with a corresponding influence on the critical temperature for superconductivity \cite{watanabeOrderedStructureTitanium1966,banusElectricalMagneticProperties1972,hulmSuperconductivityTiONbO1972a}.\\ 

\subsection*{Epitaxial growth phase diagram of the Ti-O system}

Our main experiments are summarized in Fig.\ref{Fig2}\textbf{a} where we plot the stabilized phases based on X-ray diffraction measurements in the growth temperature \Tg~versus oxygen supply \Po~parameter space. While the growth of films in ultra-high vacuum (UHV) is one of the main novelties of this work, we have explored dosing oxygen as an additional degree of freedom, as it helps accentuate how the benefits of the high temperature growth extend over a large parameter space. In this work, we quantify \Po~by  the local pressure of gas at the oxygen injection nozzle measured by a capacitive manometer (Fig.\ref{SI_Nozzle}). Inspecting the \Tg-dependent oxidation states, we found two opposing trends. The first trend canonically follows the basic knowledge of thermodynamics, that is, at relatively low \Tg~and high \Po, the oxidation states are lowered as \Tg~is increased while \Po~is constant. An Ellingham diagram predicts lower oxidation states of oxides, at higher temperatures and lower oxygen partial pressures~\cite{shangEllinghamDiagramsBinary2024}. This guideline for synthesis is practiced widely in a multitude of thin film growth recipes. We have named this behavior as the ``bulk-thermodynamic'' epitaxy regime in this article, as indicated in the upper left corner of Fig.\ref{Fig2}\textbf{a}. In contrast, we have demonstrated that it is possible to obtain a similar series of samples while closing the oxygen injection valve and maintaining the UHV condition of the chamber (Fig.\ref{SI_RGA}). Working along the bottom row of Fig.\ref{Fig2}\textbf{a}, we observed that the Ti-O films get more oxidized as the \Tg~increases (Fig.\ref{SI_UHV}). This growth mode appears to be against the thermodynamic law described above, and importantly, using \Tg~above 1000~$^\circ$C could yield phase-pure epitaxial TiO and Ti$_2$O$_3$ films while maintaining comparable growth rates (Fig.\ref{Fig2}\textbf{b}). 

For the Ti$_2$O phase, a single-phase film grew only within a narrow growth window around \Tg~= 1000~$^\circ$C and \Po~= 0.046 mbar~(Fig.\ref{SI_Ti2O}). This highlights the difficulty of stabilizing uncommon oxidation states in epitaxial thin films. Its sensitivity against the \Po variation is in contrast with that of the single-phase TiO films that can be grown at \Tg~= 1100~$^\circ$ over a range of the \Po.

\begin{figure}[t]
  \centering
   \includegraphics[width=85mm]{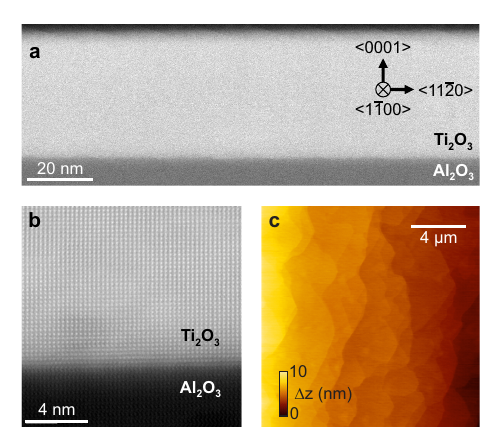}
   \caption{Microstructure analysis of a Ti$_2$O$_3$ film grown at \Tg=1300$^\circ$C in UHV in the diffusion-controlled window. \textbf{a},\textbf{b}, Wide-range TEM image (\textbf{a}) and the zoom-in (\textbf{b}) of corresponding heterointerface. \textbf{c}, Atomic force microscopy scan of the surface.}
   \label{Fig3}
\end{figure}

To account for the non-gas source oxygen supply we turn to oxygen supply from the \ALO~substrates as the leading candidate. There are two pathways which could contribute the oxygen. The first is O$_2$(\textit{g}) generated from the dissociation of \ALO(\textit{s})~at elevated temperatures. Following this hypothesis, if we rely on Ellingham chemistry to drive oxidation in the system, we would need to generate oxygen pressures far beyond those achieved with oxygen injection in Fig.\ref{Fig2} given that the temperatures are significantly elevated. However, the oxygen partial pressure inside the vacuum chamber measured using a residual gas analyzer is close to the noise limit when the substrate is at \Tg~1300$^{\circ}$ (Fig.\ref{SI_RGA}), which puts this scenario in doubt. The second pathway is oxygen diffusion from within \ALO. To check the plausibility of this hypothesis, we consider the diffusion coefficient $D$($T$) of \ALO~at elevated temperatures. Ref.\cite{heuerOxygenAluminumDiffusion2008} reports a value of approximately $D$=$10^{-22}$~cm$^2$/s of $^{18}$O at temperatures around $T$=1300~$^\circ$C. Thus, a characteristic diffusion length of angstroms on the time scale of seconds is to be expected around the deposition temperatures of our thin films, thereby realizing an effective diffusion flux fast enough to bind with impinging titanium as the growth proceeds. Therefore, we associate the corresponding growth window of the phase diagram as the diffusion-controlled growth regime (bottom right corner of Fig.\ref{Fig2}\textbf{a}).\\

\subsection*{Surface/interface characterizations of the diffusion-enabled Ti$_2$O$_3$ film}  

In addition to diffusion, the high temperatures during the growth promote all forms of kinetic processes, including adatom mobility and desorption processes. The final crystallinity of a thin film is an outcome of numerous energy- and entropy-driven kinetic processes, but in general is enhanced when surface species can settle to low-energy bonding configurations. Accordingly, the films grown in the diffusion-controlled growth regime (i.e., high \Tg~and low \Po) here have notably high quality, as evidenced in Fig.\ref{Fig3} where we present a microstructure analysis of a Ti$_2$O$_3$ film grown at \Tg=1300$^{\circ}$C in UHV. Figure \ref{Fig3}\textbf{a} illustrates the high-angle annular dark-field scanning transmission electron microscopy (HAADF-STEM) image taken from [1$\overline{1}$00] zone axis. This wide-view image confirms the absence of large defects over a long length scale and demonstrates uniform thickness throughout the film. The detailed view of Fig.\ref{Fig3}\textbf{b} shows that the structure of the film layer matches that of the Al$_2$O$_3$ substrate. In additional imaging with [11$\overline{2}$0] zone axis given in Fig.\ref{SI_TEM}, pairs of Ti atoms are visible and aligned along the $c$-axis, which is consistent with the corundum structure of Ti$_2$O$_3$. At the interface, a layer of about two unit cells is disordered, likely due to defects formed to relieve the 8\% strain mismatch between the Ti$_2$O$_3$ layer and Al$_2$O$_3$ substrate. The UHV-grown Ti$_2$O$_3$ film also reveals a smooth surface in atomic force microscopy, as shown in Fig.\ref{Fig3}\textbf{c}. The surface features atomic terraces with step bunching that appear to result from the high-temperature growth. The HAADF-STEM and atomic force microscopy images corroborate the ability of the diffusion-enabled growth to produce oxide films with well-defined interface and surface structures.\\

\begin{figure*}[t]
  \centering
   \includegraphics[width=170mm]{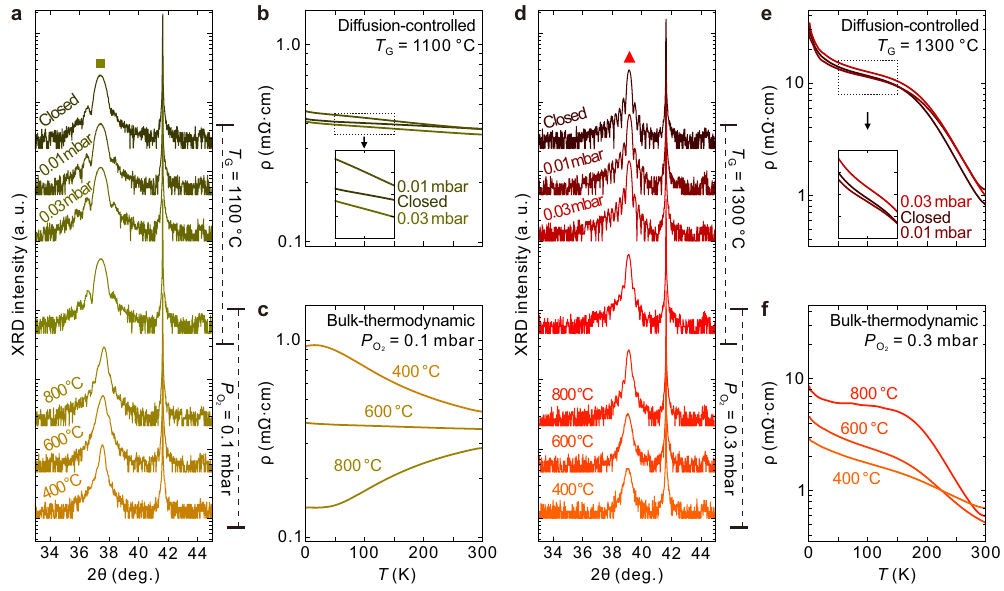}
   \caption{Comparison of films grown under conditions spanning the bulk-thermodynamic and diffusion-controlled parameter space. \textbf{a}, X-ray diffraction data of the TiO films grown under varying \Tg~and \Po~conditions. \textbf{b},\textbf{c}, Temperature dependent resistivity of TiO films grown in the diffusion-controlled regime (\textbf{b}) and the bulk-thermodynamic regime (\textbf{c}). \textbf{d}, X-ray diffraction data of the Ti$_2$O$_3$ films grown under varying \Tg~and \Po~conditions. \textbf{e},\textbf{f}, Temperature dependent resistivity of Ti$_2$O$_3$ films grown in the diffusion-controlled regime (\textbf{e}) and the bulk-thermodynamic regime (\textbf{f}).}
   \label{Fig4}
\end{figure*}

\subsection*{Comparative analysis of the bulk-thermodynamic and diffusion-controlled growths}

We now discuss the diffusion-controlled growth as an effective means of fabricating transition metal oxide films with consistent structural and electronic properties. Figure \ref{Fig4} formulates a comparative analysis of the diffusion-controlled and bulk-thermodynamic growths, focusing on the TiO structure in panels \textbf{a-c} and Ti$_2$O$_3$ in panels \textbf{d-f}. Both data sets contain a comparison of X-ray diffraction data and electrical transport and all films are phase pure as verified using asymmetric diffraction. Focusing first on the TiO structure, in the diffusion-controlled regime at high temperatures we observe a wide \Po-window at \Tg~=~$1,100\,^{\circ}{\rm C}$ where the diffraction features are high reproducible (Fig.\ref{Fig4}\textbf{a}) and electrical transport agrees qualitatively and quantitatively between samples (Fig.\ref{Fig4}\textbf{b}). In contrast, when growing at lower temperatures in the bulk-thermodynamic process, we see that the diffraction features are inferior to the high-temperature films, as inferred by the intensity of peaks and a suppression of Laue oscillations. It is also clear that the transport features of the films grown from \Tg=400$\sim$800$^\circ$C at a constant \Po~of 1$\times10^{-1}$~mbar differ not only in magnitude but also in the sign of $d\rho/dT$ (Fig.\ref{Fig4}\textbf{c}). A similar set of data is obtained for panels \textbf{d-f} for Ti$_2$O$_3$; diffraction is improved and very consistent sample-to-sample at high temperatures in the diffusion-controlled epitaxy regime, which in turn also produces samples with highly reproducible transport features and stronger insulating transport (i.e. a larger negative $d\rho/dT$) in samples grown both in UHV and with finite oxygen injected into the chamber. Similar to the TiO samples, Ti$_2$O$_3$ samples grown in the bulk-thermodynamic regime show varying electrical properties. The decrease in resistivity at low \Tg~is analogous to the low-temperature grown metallic Ti$_2$O$_3$ films from Ref.\cite{yoshimatsuMetallicGroundStates2020a}, but we didn't observe a positive $d\rho/dT$ in our study.\\

\subsection*{The diffusion-controlled growth as a self-regulated growth mechanism}

Primarily for compounds of $p$-block elements, highly crystalline thin films are often been achieved by a self-regulated growth mode often called adsorption-controlled growth, in which stoichiometric compounds are thermodynamically enforced with an oversupply of volatile elements. Its usefulness is widely recognized, for example in GaAs \cite{chungUltrahighqualityTwodimensionalElectron2021} and ZnO \cite{falsonMgZnOZnOHeterostructures2016}. While the advantages of self-regulated epitaxy are obvious, it is difficult to achieve in compounds of transition metals, especially refractory metals, where elements vaporize well above the regular temperature range of thin film growth techniques. Additionally, transition metals often have several stable oxidation states, which translates into poor thermodynamic selectivity for each. 

Addressing challenges associated with accessing adsorption-controlled growth windows of transition metal oxides is a central theme in the oxide MBE community~\cite{henslingStateArtTrends2024a}. In hybrid MBE, the low-vapor-pressure metals are delivered in the form of metal-organic precursors from both liquid and solid forms \cite{jalanGrowthHighqualitySrTiO32009,jalanMolecularBeamEpitaxy2009,brahlekFrontiersGrowthComplex2018a,nunnNovelSynthesisApproach2021a}. The volatility of the precursors is significantly higher than that of the pure metal, which enhances accessibility to adsorption-controlled growth windows. In particular, the hybrid MBE has realized epitaxial SrTiO$_3$ of the highest quality exhibiting outstanding electron mobility and strain-induced ferroelectricity~\cite{sonEpitaxialSrTiO3Films2010,haislmaierStoichiometryKeyFerroelectricity2016}. However, the adsorption-controlled growth of rare-earth titanates, where the titanium is in a 3+ oxidation state, showed a limited growth window depending on the film thickness~\cite{moetakefGrowthWindowEffect2013}. This complexity can be attributed to the abundant oxygen in the titanium-tetraisopropoxide precursor compound where the titanium is in a 4+ oxidation state. Another approach is to use suboxides as source materials which may have high volatility \cite{hoffmannEfficientSuboxideSources2020a}, for example in the case of Ga$_2$O$_3$ this technique helped to achieve an adsorption-controlled growth mode \cite{vogtAdsorptioncontrolledGrowthGa2O32021a}. We note however that the congruent evaporation of suboxide sources is available only for some elements~\cite{adkisonSuitabilityBinaryOxides2020a}. The commonality between these approaches is in providing source material that is already (at least partially) oxidized.

The ability to produce consistent films within finite growth windows (Fig.\ref{Fig4}) qualifies diffusion-controlled growth as a self-regulated growth mechanism for transition metal oxides, distinct from the two alternative MBE techniques. The diffusion-controlled growth tunes the metal-oxygen stoichiometry by the growth temperature while sticking to the elemental source of transition metal with little volatility. The diffusion-controlled epitaxy window is strongly reducing in nature due to the metallic flux and high temperatures. We therefore see our method useful in stabilizing epitaxial suboxides and perhaps in the discovery of epitaxially stabilized materials with unconventional oxidation states. In this study, we did not find the diffusion-controlled window for the Ti$^{4+}$ state (TiO$_2$) which is readily attainable by the adsorption-controlled growth with pre-oxidized sources. Thus, these two approaches can complement each other to accomplish a wide range of possible oxidation states.  \\

\begin{figure}[t]
  \centering
   \includegraphics[width=85mm]{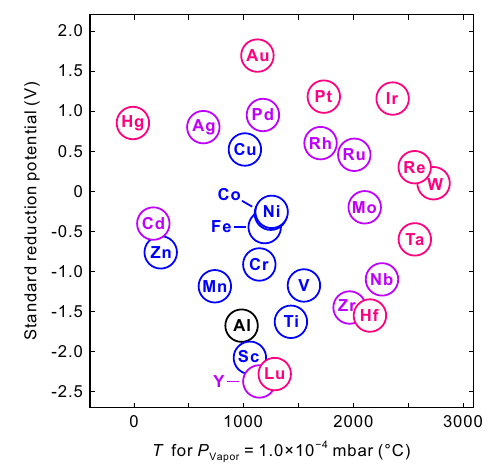}
   \caption{Reduction potential of 3$d$ (blue), 4$d$ (purple), and 5$d$ (pink) metals \cite{rumbleCRCHandbookChemistry2024} versus the temperature in Celsius required to achieve 10$^{-4}$~mbar vapor pressure.}
   \label{Fig5}
\end{figure}

\subsection*{Prospective materials suitable for the diffusion-controlled growth}

Looking forward, it will be interesting to explore the bounds of the diffusion enabled epitaxy when it comes to other transition metals from 3$d$, 4$d$ and 5$d$ rows, and also rare-earth metals. Currently, our expectation is that the approach will excel in circumstances where the reduction potential of the metal is similar to that of Al (which includes Ti). That is only half the process, however, as it is essential that the sticking coefficient of materials be sufficiently large so the film can grow even at very high temperatures without desorbing. At a given diffusion rate set by the \Tg, growth rate and time are tuning parameters for the oxidation, as shown in Fig.\ref{SI_Figtunability}. To fully capture those processes, it is essential to consider both the vapor pressure of the impinging metal, and also the various suboxides which may form on the substrate surface as oxidation passes through various states, however not all these factors are known experimentally, although there are reports available \cite{lamoreauxHighTemperatureVaporization1987} which are complemented by calculations \cite{adkisonSuitabilityBinaryOxides2020a}. We note that we have already encountered surprises related to the volatility of surface species in our data. A thickness anomaly was seen at \Tg = 1100~$^\circ$C in the series of UHV-grown samples (Fig.\ref{Fig2}\textbf{b}), which was again evident in the diffusion-controlled growths shown in Fig.\ref{Fig4}\textbf{a} and \textbf{d}. Despite being grown at higher temperatures and the Ti metal flux is identical between growths, the Ti$_2$O$_3$ phase grows notably faster than TiO.

In the absence of a clear picture of oxidation dynamics on the surface of the substrate, in Fig.\ref{Fig5} we have plotted the reduction potential of 3$d$, 4$d$ and 5$d$ metals versus the temperature required to achieve 10$^{-4}$mbar of vapor pressure. The volatility of the metal species is well known, and therefore we have chosen to use it for the analysis despite the known shortcomings discussed above. The most promising candidates for the approach are in the bottom right of the diagram and are worth prioritizing going forward.

\section*{Conclusion}
The data presented demonstrates a novel self-regulated growth mechanism of epitaxial titanium oxide films that can utilize oxygen diffusion from the substrate as the sole source of oxidation. The oxygen diffusion is triggered by high growth temperatures that have recently become technically available due to laser heating methods. The diffusing oxygen was found to be capable of compensating for insufficient supplied gaseous oxygen, letting the growing layer find its optimal form and enabling oxidation-state-specific, high-quality, and highly reproducible single phase thin films. The heteroepitaxial TiO and Ti$_2$O$_3$ films display consistent electrical characteristics over a wide window of the growth parameter space. Looking forward, we hope to broaden this approach to include 4$d$ and 5$d$ compounds where the interplay of strongly correlated electrons and strong spin-orbit interactions gives rise to a rich variety of topological phases~\cite{manzeli2DTransitionMetal2017,armitageWeylDiracSemimetals2018}. Realizing ultra-high purity films of such materials is therefore anticipated to be broadly impactful. Moreover, the approach is naturally compatible with chamber operating in UHV without oxygen sources, opening a range of opportunities to craft hybrid interfaces between reduced oxide and metal (or metalloid) layers. 

\section*{Methods}
\textit{Molecular beam epitaxy}: Films were prepared using a molecular beam epitaxy (MBE) apparatus with a base pressure below 1$\times$10$^{-9}$~mbar. The primary pump on the chamber is a maglev turbo molecular pump. Films were grown using a home-built CO$_2$ laser heating apparatus capable of heating substrates in excess of 2000~$^\circ$C based on a Coherent E-1000 laser. Prior to growth, the substrates were annealed at $T$=1550~$^\circ$C for 10 minutes to obtain atomically flat terraces. Thermometry is performed coaxially with the laser using a long wavelength pyrometer observing the wafer from the back side. Films were synthesized at a constant temperature, requiring real-time feedback to the laser power output. Ti flux was provided from a high temperature effusion cell with a titanium-tolerant crucible operating up to 1950~$^\circ$C. Flux was measured before and after growth using a quartz crystal microbalance and was kept at 0.09 \AA/s unless otherwise indicated. The growth chamber pressure value can be affected by elemental evaporation, degassing, and shutter motions. To ensure a constant oxygen flow rate, the pressure behind the gas delivery nozzle was separately measured and maintained. The idle state relationship between the growth chamber pressure and the nozzle pressure (\Po) is given in the Fig.\ref{SI_Nozzle}. Oxygen was provided through a set of two variable leak valves in series with a barometer in between. The pressure in the intermediate volume was used as a tunable parameter. The leak valve between the volume of pipes containing the barometer and the growth chamber is at a set impedance, while the leak valve between the barometer and oxygen cylinder is controlled variably throughout the growth process. Oxygen is delivered through a water cooled tube to within a few centimeters of the substrate. The chamber pressure, partial pressure of oxygen resolved in the residual gas analyzer, and baratron pressure were all used to inform the synthesis and ensure reproducibility. All films were found to be stable after extracting them from the MBE, requiring no additional capping layers. 

\textit{Electron microscopy}: A focused ion beam (FEI Helios NanoLab 600 DualBeam) was used to prepare the cross-section samples for the S/TEM analyses. All samples were initially thinned with 5 kV Ga ions and then polished at 2 kV to minimize surface damage. We used the Thermo Scientific Themis Z S/TEM with a 64-200 mrad HAADF detector to image the Ti$_2$O$_3$ film grown on Sapphire. The microscope operated at 300 kV with a semi-convergence angle of 30 mrad. To improve the signal-to-noise ratio, we captured HAADF-STEM images as a series of 20 fast scan images (2048 × 2048 pixels, 200 ns per frame) and then averaged them.

\textit{X-ray diffraction}: X-ray diffraction data was obtained using a Rigaku Smartlab diffractometer using a 2-bounce Ge (220) monochromator and Cu-$K\alpha1$ radiation. Asymmetric diffraction data was obtained using the 1D mode of the Rigaku HyPix 3000 detector. 

\textit{Electrical transport}: Transport data was obtained used a Quantum Design Dynacool operating down to 1.7~K. Samples were probed in a four-point geometry using Al wire as a contact material. Resistance data was obtained using the internal resistance bridge of the Dynacool system.

\section*{Acknowledgments}
We acknowledge funding provided by the Gordon and Betty Moore Foundation’s EPiQS Initiative (Grant number GBMF10638), and the Institute for Quantum Information and Matter, a NSF Physics Frontiers Center (NSF Grant PHY-1733907). We acknowledge the Beckman Institute for their support of the X-Ray Crystallography Facility at Caltech. Electron microscopy was performed at the Center for Electron Microscopy and Analysis (CEMAS) at The Ohio State University.

\clearpage

\section*{}
\bibliography{ref.bib}

\renewcommand{\thefigure}{S\arabic{figure}}

\setcounter{figure}{0}

\begin{table*}[h]
\centering
\caption{
	\label{table} 
	Structural data and epitaxial relationships
      }
        \begin{tabular}{c|c|cc}\hline\hline
          Compounds & Space group &  ~~~~~Out-of-plane~~~~~ & ~~~~~In-plane~~~~~ \\\hline
          Al$_2$O$_3$ & $R\overline{3}c$ (167) & [0001] & [1$\overline{1}$00]\\\hline
          Ti & $P6_3/mmc$ (194) & [0001]  &  [11$\overline{2}$0]\\
          Ti$_2$O & $P\overline{3}m1$ (164) &  [0001] & [11$\overline{2}$0]\\
          TiO & $Fm\overline{3}m$ (225) &  [111] & [1$\overline{1}$0]\\
          Ti$_2$O$_3$ & $R\overline{3}c$ (167)  &  [0001] & [1$\overline{1}$00]\\
    \hline\hline
  \end{tabular}
\end{table*}

\begin{figure*}[h]
  \centering
   \includegraphics[width=127.5mm]{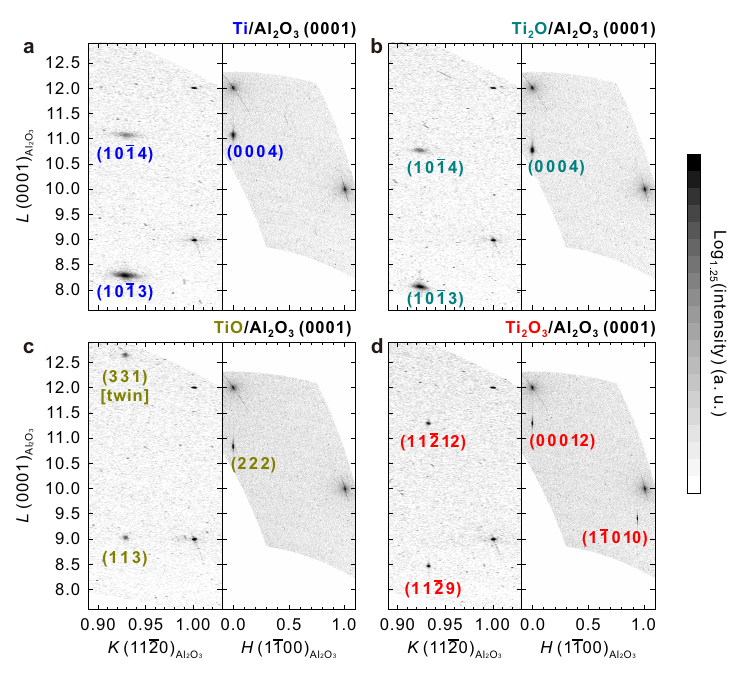}
   \caption{\textbf{a}-\textbf{d}, XRD reciprocal space mappings of representative Ti (\textbf{a}), Ti$_2$O (\textbf{b}), TiO (\textbf{c}), and Ti$_2$O$_3$ (\textbf{d}) samples. Measurements were conducted along Al$_2$O$_3$ (11$\overline{2}$0) and (1$\overline{1}$00) in-plane orientations.}
   \label{SI_RSM}
\end{figure*}

\begin{figure*}[h]
  \centering
   \includegraphics[width=85mm]{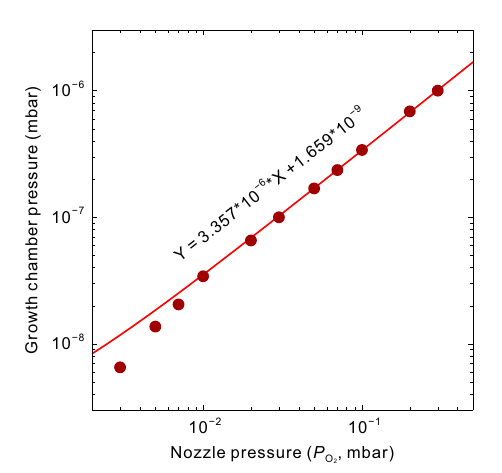}
   \caption{Calibration of the idle state relationship between growth chamber pressure and nozzle pressure.}
   \label{SI_Nozzle}
\end{figure*}

\begin{figure*}[h]
  \centering
   \includegraphics[width=170mm]{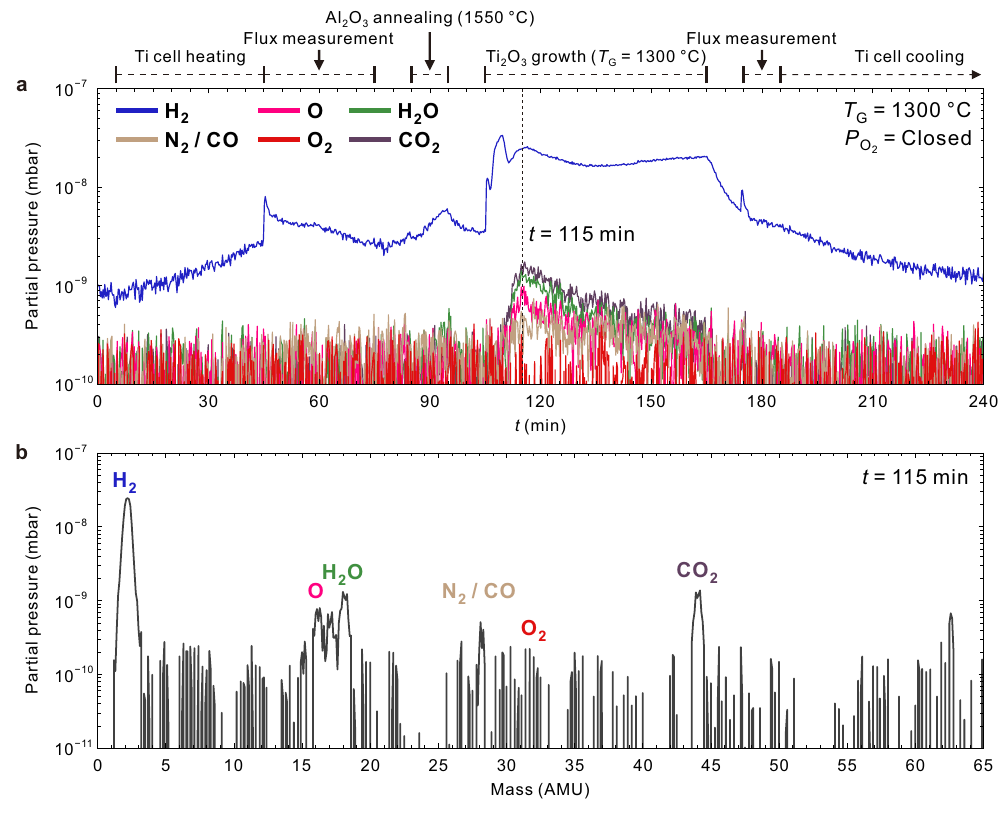}
   \caption{Residual gas analyzer (RGA) measurements for the Ti$_2$O$_3$/Al$_2$O$_3$ (0001) growth at \Tg~= 1300 $^\circ$C without injecting oxygen. \textbf{a}, Time-dependent partial pressures of H$_2$, O, H$_2$O, N$_2$ or CO, O$_2$, and CO$_2$ gas molecules. \textbf{b}m, RGA scan at \textit{t}~= 115 min.}
   \label{SI_RGA}
\end{figure*}

\begin{figure*}[h]
  \centering
   \includegraphics[width=85mm]{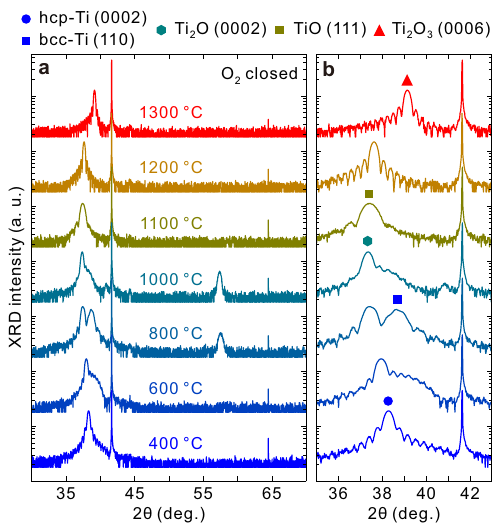}
   \caption{The XRD of Ti-O films grown at UHV. \textbf{a},\textbf{b}, Wide range XRD scans (\textbf{a}) and zooms in around the \ALO~(0006) peak (\textbf{b})}
   \label{SI_UHV}
\end{figure*}

\begin{figure*}[h]
  \centering
   \includegraphics[width=85mm]{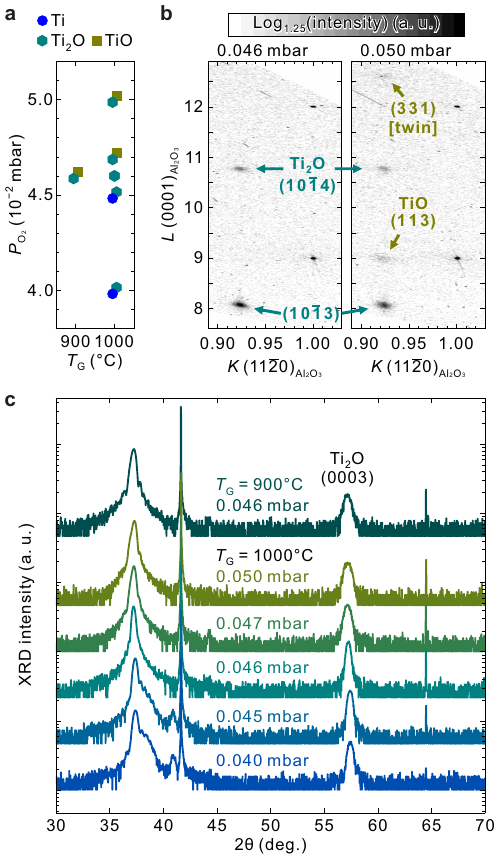}
   \caption{Growth optimization of a single-phase Ti$_2$O film. \textbf{a}, Narrow-range growth phase diagram in the $T$-$P_\mathrm{O_2}$ parameter space. \textbf{b}, Reciprocal space mappings of the single Ti$_2$O phase and Ti$_2$O-TiO mixed phase samples measured along Al$_2$O$_3$ (11$\overline{2}$0) orientation. The single (mixed) phase film was grown at \Tg = 1000~$^\circ$C and \Po = 0.046 mbar (0.050 mbar) condition. \textbf{c}, XRD of films presented in \textbf{a}}
   \label{SI_Ti2O}
\end{figure*}

\begin{figure*}[h]
  \centering
   \includegraphics[width=170mm]{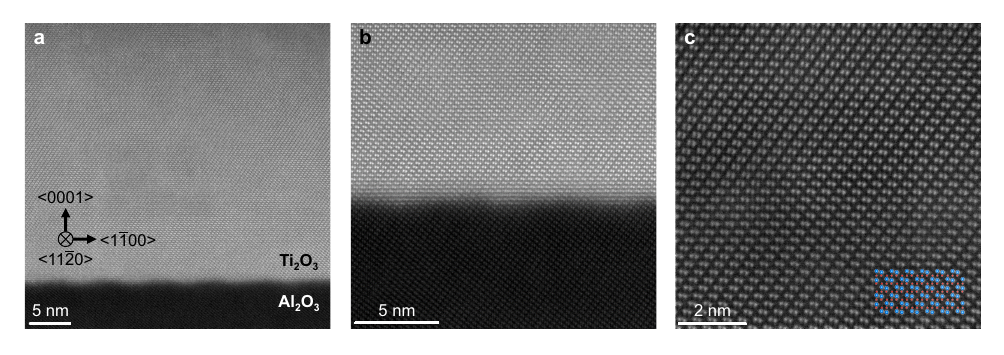}
   \caption{Transmission electron microscopy analysis of a \Tg=1300$^\circ$C, UHV-grown film along the [11$\overline{2}$0] direction (orthogonal to the data shown in the main text). \textbf{a},\textbf{b}, Wide view (\textbf{a}) and close up (\textbf{b}) of the interface. Corrugations at the interface are likely associated with the relaxation of strain along the surface owing to the $\sim$8\% lattice mismatch between layers. \textbf{c}, Close up of the Ti$_2$O$_3$ film with atoms overlaid.}
   \label{SI_TEM}
\end{figure*}

\begin{figure*}[h]
  \centering
   \includegraphics[width=85mm]{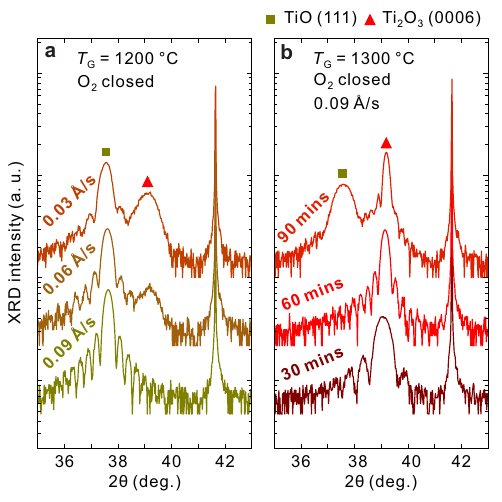}
   \caption{Tunability of diffusion-controlled oxidation of the Ti-O system. \textbf{a}, XRD of films when the Ti flux
   is varied at \Tg~= 1200 $^\circ$C. The growth times were 60 mins, 90 mins, and 180 mins for the 0.09 \AA/s, 0.06 \AA/s, and 0.03 \AA/s growths, to keep the amount of evaporated titanium the same. \textbf{b}, XRD of films when the growth time is varied at \Tg~= 1300 $^\circ$C. 
   Slower growth can promote more oxidation of films through sufficient time for diffusion. The limited oxidation of the 90 mins sample in \textbf{b} is presumably because oxygen is diffusing slowly through Ti$_2$O$_3$.}
   \label{SI_Figtunability}
\end{figure*}


{\setstretch{0.75}

\begin{turnpage}\begin{table}[h]
\begin{tabular}{ccccc}
\hline
\multicolumn{1}{|c|}{\textbf{Sc}}                       & \multicolumn{1}{c|}{\textbf{Ti}}                       & \multicolumn{1}{c|}{\textbf{V}}                        & \multicolumn{1}{c|}{\textbf{Cr}}                      & \multicolumn{1}{c|}{\textbf{Mn}}                       \\ \hline
\multicolumn{1}{|c|}{1.08E+03}                          & \multicolumn{1}{c|}{1.43E+03}                          & \multicolumn{1}{c|}{1.54E+03}                          & \multicolumn{1}{c|}{1.09E+03}                         & \multicolumn{1}{c|}{775}                               \\ \hline
\multicolumn{1}{|c|}{–2.077}                            & \multicolumn{1}{c|}{–1.628}                            & \multicolumn{1}{c|}{–1.175}                            & \multicolumn{1}{c|}{-0.913}                           & \multicolumn{1}{c|}{-1.185}                            \\ \hline
\multicolumn{1}{|c|}{(Sc3+) + 3(e1-) =\textgreater Sc}  & \multicolumn{1}{c|}{(Ti2+) + 2(e1-) =\textgreater Ti}  & \multicolumn{1}{c|}{(V2+) + 2(e1-) =\textgreater V}    & \multicolumn{1}{c|}{(Cr2+) + 2(e1-) =\textgreater Cr} & \multicolumn{1}{c|}{(Mn2+) + 2(e1-) =\textgreater Mn}  \\ \hline
\multicolumn{1}{|c|}{\textbf{Y}}                        & \multicolumn{1}{c|}{\textbf{Zr}}                       & \multicolumn{1}{c|}{\textbf{Nb}}                       & \multicolumn{1}{c|}{\textbf{Mo}}                      & \multicolumn{1}{c|}{\textbf{Tc}}                       \\ \hline
\multicolumn{1}{|c|}{1.33E+03}                          & \multicolumn{1}{c|}{1.92E+03}                          & \multicolumn{1}{c|}{2.25E+03}                          & \multicolumn{1}{c|}{2.08E+03}                         & \multicolumn{1}{c|}{X Unstable}                        \\ \hline
\multicolumn{1}{|c|}{–2.372}                            & \multicolumn{1}{c|}{–1.45}                             & \multicolumn{1}{c|}{-1.099}                            & \multicolumn{1}{c|}{-0.2}                             & \multicolumn{1}{c|}{0.4}                               \\ \hline
\multicolumn{1}{|c|}{(Y3+) + 3(e1-)  =\textgreater Y}   & \multicolumn{1}{c|}{(Zr4+) + 4(e1-)  =\textgreater Zr} & \multicolumn{1}{c|}{(Nb3+) + 3(e1-) =\textgreater Nb}  & \multicolumn{1}{c|}{(Mo3+) + 3(e1-) =\textgreater Mo} & \multicolumn{1}{c|}{(Tc2+) + 2(e1-) =\textgreater Tc}  \\ \hline
\multicolumn{1}{|c|}{\textbf{Lu}}                       & \multicolumn{1}{c|}{\textbf{Hf}}                       & \multicolumn{1}{c|}{\textbf{Ta}}                       & \multicolumn{1}{c|}{\textbf{W}}                       & \multicolumn{1}{c|}{\textbf{Re}}                       \\ \hline
\multicolumn{1}{|c|}{1.35E+03}                          & \multicolumn{1}{c|}{2.02E+03}                          & \multicolumn{1}{c|}{2.56E+03}                          & \multicolumn{1}{c|}{2.77E+03}                         & \multicolumn{1}{c|}{2.53E+03}                          \\ \hline
\multicolumn{1}{|c|}{–2.28}                             & \multicolumn{1}{c|}{–1.55}                             & \multicolumn{1}{c|}{–0.6}                              & \multicolumn{1}{c|}{0.1}                              & \multicolumn{1}{c|}{0.3}                               \\ \hline
\multicolumn{1}{|c|}{(Lu3+) + 3(e1-)  =\textgreater Lu} & \multicolumn{1}{c|}{(Hf4+) + 4(e1-)  =\textgreater Hf} & \multicolumn{1}{c|}{(Ta3+) + 3(e1-) =\textgreater Ta}  & \multicolumn{1}{c|}{(W3+) + 3(e1-) =\textgreater W}   & \multicolumn{1}{c|}{(Re3+) + 3(e1-) =\textgreater Re}  \\ \hline
                                                        &                                                        &                                                        &                                                       &                                                        \\ \hline
\multicolumn{1}{|c|}{\textbf{Fe}}                       & \multicolumn{1}{c|}{\textbf{Co}}                       & \multicolumn{1}{c|}{\textbf{Ni}}                       & \multicolumn{1}{c|}{\textbf{Cu}}                      & \multicolumn{1}{c|}{\textbf{Zn}}                       \\ \hline
\multicolumn{1}{|c|}{1.13E+03}                          & \multicolumn{1}{c|}{1.26E+03}                          & \multicolumn{1}{c|}{1.26E+03}                          & \multicolumn{1}{c|}{1.01E+03}                         & \multicolumn{1}{c|}{244}                               \\ \hline
\multicolumn{1}{|c|}{-0.447}                            & \multicolumn{1}{c|}{-0.28}                             & \multicolumn{1}{c|}{-0.257}                            & \multicolumn{1}{c|}{0.521}                            & \multicolumn{1}{c|}{–0.7618}                           \\ \hline
\multicolumn{1}{|c|}{(Fe2+) + 2(e1-) =\textgreater Fe}  & \multicolumn{1}{c|}{(Co2+) + 2(e1-) =\textgreater Co}  & \multicolumn{1}{c|}{(Ni2+) + 2(e1-) =\textgreater Ni}  & \multicolumn{1}{c|}{(Cu+) + (e1-) =\textgreater Cu}   & \multicolumn{1}{c|}{(Zn2+) + 2(e1-)  =\textgreater Zn} \\ \hline
\multicolumn{1}{|c|}{\textbf{Ru}}                       & \multicolumn{1}{c|}{\textbf{Rh}}                       & \multicolumn{1}{c|}{\textbf{Pd}}                       & \multicolumn{1}{c|}{\textbf{Ag}}                      & \multicolumn{1}{c|}{\textbf{Cd}}                       \\ \hline
\multicolumn{1}{|c|}{1.98E+03}                          & \multicolumn{1}{c|}{1.68E+03}                          & \multicolumn{1}{c|}{1.19E+03}                          & \multicolumn{1}{c|}{811}                              & \multicolumn{1}{c|}{175}                               \\ \hline
\multicolumn{1}{|c|}{0.455}                             & \multicolumn{1}{c|}{0.6}                               & \multicolumn{1}{c|}{0.951}                             & \multicolumn{1}{c|}{0.7996}                           & \multicolumn{1}{c|}{–0.4030}                           \\ \hline
\multicolumn{1}{|c|}{(Ru2+) + 2(e1-) =\textgreater Ru}  & \multicolumn{1}{c|}{(Rh+) + (e1-) =\textgreater Rh}    & \multicolumn{1}{c|}{(Pd2+) + 2(e1-)  =\textgreater Pd} & \multicolumn{1}{c|}{(Ag1+) + (e1-) =\textgreater Ag}  & \multicolumn{1}{c|}{(Cd2+) + 2(e1-)  =\textgreater Cd} \\ \hline
\multicolumn{1}{|c|}{\textbf{Os}}                       & \multicolumn{1}{c|}{\textbf{Ir}}                       & \multicolumn{1}{c|}{\textbf{Pt}}                       & \multicolumn{1}{c|}{\textbf{Au}}                      & \multicolumn{1}{c|}{\textbf{Hg}}                       \\ \hline
\multicolumn{1}{|c|}{X Toxic}                           & \multicolumn{1}{c|}{2.36E+03}                          & \multicolumn{1}{c|}{1.62E+03}                          & \multicolumn{1}{c|}{1.13E+03}                         & \multicolumn{1}{c|}{-8.9}                              \\ \hline
\multicolumn{1}{|c|}{}                                  & \multicolumn{1}{c|}{1.156}                             & \multicolumn{1}{c|}{1.18}                              & \multicolumn{1}{c|}{1.692}                            & \multicolumn{1}{c|}{0.851}                             \\ \hline
\multicolumn{1}{|c|}{}                                  & \multicolumn{1}{c|}{(Ir3+) + 3(e1-)  =\textgreater Ir} & \multicolumn{1}{c|}{(Pt2+) + 2(e1-)  =\textgreater Pt} & \multicolumn{1}{c|}{(Au1+) + (e1-) =\textgreater Au}  & \multicolumn{1}{c|}{(Hg2+) + 2(e1-)  =\textgreater Hg} \\ \hline
                                                        &                                                        &                                                        &                                                       &                                                        \\ \cline{1-1}
\multicolumn{1}{|c|}{\textbf{Al}}                       &                                                        &                                                        &                                                       &                                                        \\ \cline{1-1}
\multicolumn{1}{|c|}{982}                               &                                                        &                                                        &                                                       &                                                        \\ \cline{1-1}
\multicolumn{1}{|c|}{–1.676}                            &                                                        &                                                        &                                                       &                                                        \\ \cline{1-1}
\multicolumn{1}{|c|}{(Al3+) + 3(e1-)  =\textgreater Al} &                                                        &                                                        &                                                       &                                                        \\ \cline{1-1}
\end{tabular}
\end{table}
\end{turnpage}
}

\clearpage
TABLE II.Table of transition metal elements and aluminum. The first entry for each element corresponds to the temperature in Celsius required to achieve 10$^{-4}$~mbar vapor pressure. The second entry is the reduction potential for the reaction between the metal and the lowest oxidation state. The third entry indicates the corresponding reduction reaction. Data are taken from Ref.\cite{rumbleCRCHandbookChemistry2024}.
\end{document}